\documentclass[english]{sbrt}
\usepackage[english]{babel}
\usepackage[utf8]{inputenc}

\usepackage{amsmath,amssymb,amsfonts,bm,graphicx}
\usepackage{algorithm} 
\usepackage{algpseudocode}
\usepackage{xcolor}
\usepackage{algorithm}
\usepackage{algpseudocode}
\algrenewcommand\algorithmicrequire{\textbf{Require:}}
\algrenewcommand\algorithmicensure{\textbf{Return:}}

\begin{document}

\title{Low-Complexity Tensor-Based Channel Parameter Estimation for mmWave MIMO-Assisted  Systems}

\author{Francisco I. T. Pires, Jean C. S. Ferreira, and Fazal-E Asim
\thanks{Francisco I. T. Pires, Jean C. S. Ferreira, and Fazal-E Asim, Department of Teleinformatics Engineering, Federal University of Ceará, Fortaleza-CE, e-mail: \{igorpires,jeancarlos8249\}@alu.ufc.br; e-mail: fazalasim@ufc.br.} %
}

\maketitle

\markboth{XLIV BRAZILIAN SYMPOSIUM ON TELECOMMUNICATIONS AND SIGNAL PROCESSING - SBrT 2026, SEPTEMBER 29TH TO OCTOBER 2ND, 2026, SALVADOR, BA}{}

\begin{abstract}
This paper presents a low-complexity tensor-based channel estimation framework for downlink mmWave massive MIMO systems using passive Butler-matrix beamforming. By modeling received signals as a third-order PARAFAC tensor, we exploit known pilot structures via bilinear alternating least-squares (BALS) optimization to reduce computational overhead. The decoupled spatial factors are then processed via 1D and 2D ESPRIT to extract angles of arrival and departure with automatic pairing. Additionally, we derive the Cramér--Rao lower bound (CRLB) for this architecture. Simulations confirm the proposed estimator outperforms conventional least-squares, achieving accurate channel reconstruction and approaching the CRLB at moderate-to-high signal-to-noise ratios.
\end{abstract}


\section{Introduction}

Millimeter-wave (mmWave) communications offer massive spectral bandwidths for 5G and future 6G systems, though they suffer from high path loss, severe attenuation, and blockage sensitivity \cite{elbir2022}. While massive multiple-input multiple-output (MIMO) arrays can mitigate these impairments, fully digital beamforming architectures require a dedicated radio-frequency (RF) chain per antenna, leading to prohibitive hardware complexity, cost, and power consumption \cite{elbir2022}. Hybrid analog/digital architectures offer a practical trade-off between spectral efficiency and hardware complexity. However, conventional hybrid structures rely on high-resolution adaptive phase shifters (PSs) for flexible beam control; these networks often incur significant insertion losses, high power overhead, and complex calibration requirements \cite{asim2023twc}. As a low-complexity, energy-efficient alternative, passive Butler matrices (BMs) can generate fixed orthogonal beams via analog discrete Fourier transform (DFT) operations, drastically reducing hardware constraints while retaining robust beamspace channel probing capabilities \cite{asim2021twc, asim2023twc}. Nevertheless, accurately isolating angles of arrival (AoAs), azimuth/elevation angles of departure (AoDs), and path gains in BM-assisted systems remains challenging due to the multidimensional coupling inherent to beamspace probing.

Tensor decompositions, particularly the parallel factor analysis (PARAFAC) model, are powerful tools for preserving uniqueness and capturing structural integrity in multidimensional signal processing \cite{sidiropoulos2017, dealmeida2023overview}. Leveraging multilinear models enables the efficient decoupling of spatial signatures, pilot domains, and fading gains under mild identifiability conditions across various millimeter-wave and integrated sensing applications \cite{liu2025nearfield}. In this paper, we propose a low-complexity tensor-based framework for joint channel parameter estimation in passive BM-assisted mmWave massive MIMO systems. The received training block is formulated as a third-order PARAFAC tensor that spans the transmit beamspace, receive spatial response, and pilot dimensions. This framework bypasses exhaustive beam-sweeping and adaptive PS calibration, recovering physical channel attributes directly from the fixed orthogonal beams. 

The main contributions of this work are summarized as follows. First, we establish a structured tensor signal model for downlink BM-assisted mmWave MIMO systems featuring uniform rectangular transmit arrays (URAs) and uniform linear receive arrays (ULAs). Second, we design a robust training stage employing constant-amplitude zero-autocorrelation (CAZAC) Zadoff--Chu sequences \cite{chu1972, andrews2022} to improve factor identifiability and noise resilience. Third, by exploiting the deterministic, known pilot matrix, we reformulate the unconstrained trilinear problem into a highly efficient bilinear alternating least-squares (BALS) optimization scheme. Fourth, we apply one- and two-dimensional ESPRIT algorithms \cite{roy1989} to the estimated spatial factors to extract paired AoAs, azimuth/elevation AoDs, and complex gains automatically. Finally, we derive the Cramér--Rao Lower Bound (CRLB) \cite{kay1993} tailored to the tensor-structured BM system as a theoretical benchmark and validate it against conventional least-squares baselines via extensive simulations.

\section{System Model}
\label{sec:sectionTwo}
We consider a downlink multiple-input multiple-output (MIMO) system in which the transmitter ($tx$) employs a Butler-matrix-assisted hybrid beamforming architecture to transmit signals to a single receiver ($rx$). The transmitter uses a reduced number of RF chains, denoted by $N_{RF}$, to feed a larger antenna aperture through passive beamforming networks. For the considered transmit aperture, let $M=M_hM_v$ denote the number of antenna elements addressed by the two-dimensional Butler beam codebook, where $M_h$ and $M_v$ are the numbers of horizontal and vertical elements, respectively.

\begin{figure}[t]
    \centering
    \includegraphics[width=0.95\linewidth]{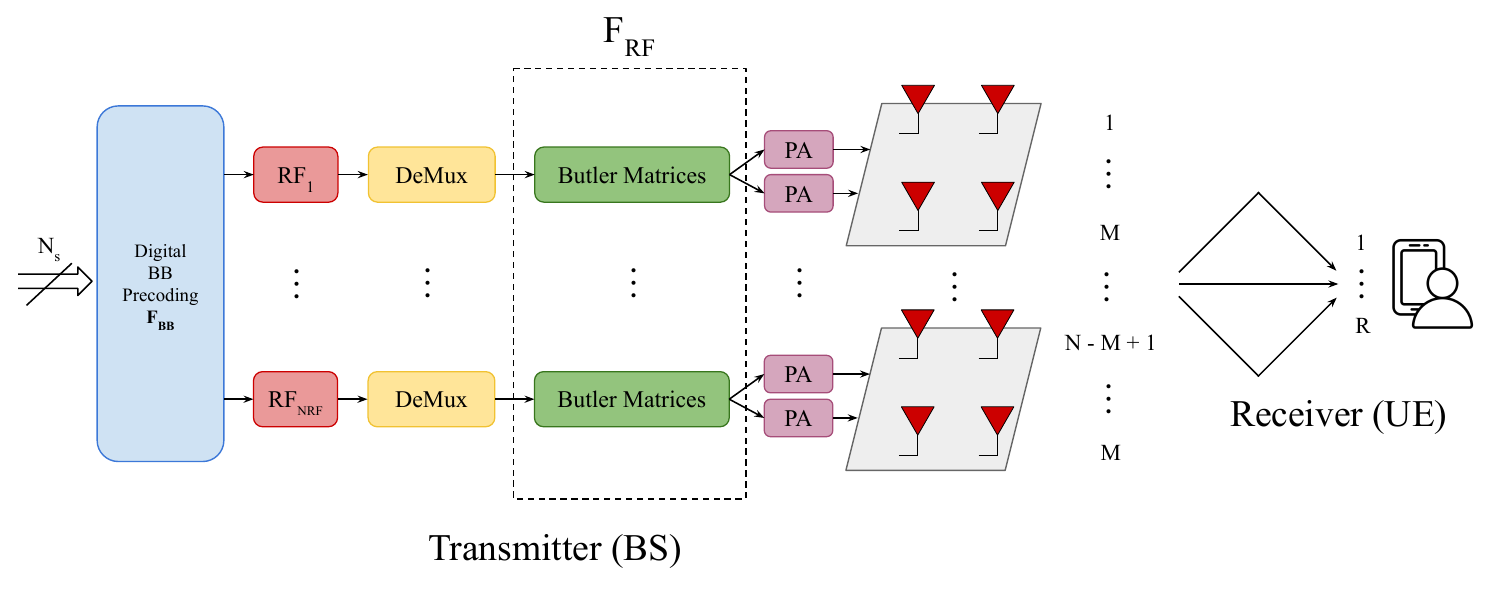}
    \caption{Butler-matrix-assisted hybrid beamforming architecture.}
    \label{fig:bm}
\end{figure}

The receiver is equipped with a fully digital ULA composed of $Q$ antenna elements \cite{asim2021twc,dealmeida2023overview}. The received signal at the receiver using the $m$-th transmit beamforming vector is given as
\begin{equation}
\label{eq:receivedSignal}
\begin{split}
& \bm{Y}_{m}=\sqrt{P_{T}}\sum_{r=1}^{R}\alpha_{r}\bm{W}^{H}\bm{a}(\phi_{r}^{rx})\bm{a}^{T}(\phi_{r}^{tx},\theta_{r}^{tx})\bm{f}_{m}^{*}(\delta_{p},\nu_{q})\bm{c}_r^{T} \\ 
&+ \bm{W}^{H}[\bm{n}_{1,m},\bm{n}_{2,m},...,\bm{n}_{L,m}]\in\mathbb{C}^{Q\times L},
\end{split}
\end{equation}
where $m=1,\ldots,M$ indexes the transmit beamforming vectors, $P_{T}$ is the instantaneous transmit power, $R$ is the number of propagation paths, and $\alpha_{r}$ denotes the complex path gain of the $r$-th path. The vectors $\bm{a}(\phi_{r}^{rx})\in\mathbb{C}^{Q\times1}$ and $\bm{a}(\phi_{r}^{tx},\theta_{r}^{tx})\in\mathbb{C}^{M\times1}$ are the receive and transmit steering vectors, respectively, $\bm{W}\in\mathbb{C}^{Q\times Q}$ denotes the receive DFT beamforming matrix, $\bm{f}_{m}(\delta_{p},\nu_{q})\in\mathbb{C}^{M\times1}$ is the $m$-th transmit beamforming vector, and $\bm{c}_{r}\in\mathbb{C}^{L\times1}$ is the pilot sequence associated with the $r$-th path. The noise vector $\bm{n}_{l,m}\sim\mathcal{CN}(\bm{0}_{Q\times1},\sigma_{n}^{2}\bm{I}_{Q})$, $l=1,\ldots,L$, represents the additive white Gaussian noise at the $l$-th pilot symbol for the $m$-th transmit beam.

We assume that the considered transmitter aperture is a uniform rectangular array (URA) in the $y-z$ plane with $M_{h}$ horizontal elements and $M_{v}$ vertical elements.

Therefore, the transmit array response to the multipath of receiver can be modeled as
\begin{equation}
\label{eq:transmitArrayResponse}
[\bm{a}(\phi_{r}^{tx},\theta_{r}^{tx})]_{m}=e^{-j\pi[(m_{h}-1)\sin\theta_{r}^{tx}\sin\phi_{r}^{tx}+(m_{v}-1)\cos\theta_{r}^{tx}]},
\end{equation}
where $\phi_{r}^{tx}$ and $\theta_{r}^{tx}$ denote the azimuth and elevation AoDs, respectively, and $m = m_{v}+(m_{h}-1)M_{v}$ with $m_{h}\in\{1,\ldots,M_{h}\}$ and $m_{v}\in\{1,\ldots,M_{v}\}$.

Similarly, we assume that each receiver is equipped with a uniform linear array (ULA). The receive steering vector $\bm{a}(\phi_{r}^{rx})\in\mathbb{C}^{Q\times1}$ is defined with the angle of arrival (AoA) $\phi_{r}^{rx}$
\begin{equation}
\bm{a}(\phi_{r}^{rx})=[1,e^{-j\frac{2\pi}{\lambda}d_{r}\sin(\phi_{r}^{rx})},...,e^{-j\frac{2\pi}{\lambda}(Q-1)d_{r}\sin(\phi_{r}^{rx})}]^{T}.
\end{equation}

The azimuth and elevation angles are assumed to be distributed within the angular region visible to the corresponding transmit and receive arrays.
Assuming the receive antenna array inter-element spacing $d_{r}=\frac{\lambda}{2}$ and defining the receive spatial frequency $\rho_{r}=\pi~\sin(\phi_{r}^{rx})$, the receive steering vector can be rewritten as
\begin{equation}
\bm{a}(\rho_{r})=[1,e^{-j\rho_{r}},...,e^{-j(Q-1)\rho_{r}}]^{T}\in\mathbb{C}^{Q\times1}.
\end{equation}

Now, let us consider the two-dimensional steering vector $\bm{a}(\phi_{r}^{tx},\theta_{r}^{tx})$ given in (\ref{eq:transmitArrayResponse}) and define the spatial frequencies as $\mu_{r}=\pi\sin\theta_{r}^{tx}\sin\phi_{r}^{tx}$ and $\psi_{r}=\pi\cos\theta_{r}^{tx}$ respectively.
Therefore, the steering vector can be written as the Kronecker product of the following steering vectors
\begin{equation}
\bm{a}(\mu_{r},\psi_{r})=\bm{a}_{h}(\mu_{r})\otimes \bm{a}_{v}(\psi_{r})\in\mathbb{C}^{M\times1},
\end{equation}
where
\begin{equation}
\bm{a}_{h}(\mu_{r})=[1,e^{-j\mu_{r}},...,e^{-j(M_{h}-1)\mu_{r}}]^{T}\in\mathbb{C}^{M_{h}\times1},
\end{equation}
\begin{equation}
\bm{a}_{v}(\psi_{r})=[1,e^{-j\psi_{r}},...,e^{-j(M_{v}-1)\psi_{r}}]^{T}\in\mathbb{C}^{M_{v}\times1}.
\end{equation}

The $m$-th two-dimensional beamforming vector, formed by using a network of Butler Matrices (BMs) \cite{asim2021twc,asim2023twc} to excite a URA at the transmitter, is expressed as the Kronecker product of the columns of DFT matrices as
\begin{equation}
\bm{f}_{m}(\delta_{p},\nu_{q})=\bm{f}_{h}(\delta_{p})\otimes \bm{f}_{v}(\nu_{q})\in\mathbb{C}^{M\times1},
\end{equation}
where $m=1,\ldots,M=M_{h}M_{v}$, with each value of $m$ corresponding to one pair $(p,q)$ of horizontal and vertical DFT beams. Here, $\bm{f}_{m}(\delta_{p},\nu_{q})$ is the resultant beamforming vector for probing the channel, $\delta_{p}$ is the DFT spatial frequency controlling the horizontal antenna elements, and $\bm{f}_{h}(\delta_{p})$ is defined as
\begin{equation}
\bm{f}_{h}(\delta_{p})=\frac{1}{\sqrt{M_{h}}}[1,e^{-j\delta_{p}},...,e^{-j(M_{h}-1)\delta_{p}}]^{T}\in\mathbb{C}^{M_{h}\times1},
\end{equation}
where $\delta_p = \frac{2\pi}{M_h}p$, $p = 0,\ldots, M_h - 1$. The parameter $\nu_{q}$ is the DFT spatial frequency exciting the vertical antenna elements in the URA, while $\bm{f}_{v}(\nu_{q})$ is defined as
\begin{equation}
\bm{f}_{v}(\nu_{q})=\frac{1}{\sqrt{M_{v}}}[1,e^{-j\nu_{q}},...,e^{-j(M_{v}-1)\nu_{q}}]^{T}\in\mathbb{C}^{M_{v}\times1},
\end{equation}
where $\nu_q = \frac{2\pi}{M_v}q$, $q=0,\ldots,M_{v}-1$.

To ensure tensor factor identifiability and mitigate multipath interference, we employ constant-amplitude zero-autocorrelation (CAZAC) Zadoff-Chu pilots \cite{chu1972, andrews2022}, following the design in \cite{asim2021twc}. Their constant-envelope and ideal periodic autocorrelation properties minimize the transmitter's peak-to-average power ratio (PAPR). Crucially, assigning distinct cyclic shifts of a base sequence to different propagation paths generates nearly orthogonal pilot streams. This deterministic shift design guarantees a full-column rank pilot matrix, thereby securing the uniqueness and robust separability of the latent spatial signatures during the bilinear optimization stage.
The pilot matrix is, therefore, written as
\begin{equation}
{\bm{C}} =
\begin{bmatrix}
\bm{c}_1^T\\
\bm{c}_2^T\\
\vdots\\
\bm{c}_R^T
\end{bmatrix}
\in\mathbb{C}^{R\times L}.
\end{equation}

By rearranging the scalar inner product
$\bm{a}^{T}(\phi_r^{tx},\theta_r^{tx})\bm{f}_m^*$
and defining the filtered noise matrix
$\bm{N}_m=\bm{W}^H[\bm{n}_{1,m},\dots,\bm{n}_{L,m}]$,
the received signal becomes

\begin{equation}
\label{eq:reorganizedSignal}
\begin{split}
    \bm{Y}_{m} &= \sqrt{P_{T}}\sum_{r=1}^{R}\left(\bm{f}_{m}^{H}(\delta_{p},\nu_{q})\bm{a}(\mu_{r},\psi_{r})\right)
    \\&
    \left(\bm{W}^{H}\bm{a}(\rho_{r})\right)\alpha_{r}\bm{c}_{r}^{T} + \bm{N}_{m},
\end{split}
\end{equation}
This algebraic rearrangement effectively decouples the transmitter and receiver spatial signatures, establishing the basis for the multi-dimensional tensor modeling.

To express the signal in a compact matrix form, we define the transmit array response matrix $\bm{A}_{T}=[\bm{a}(\mu_{1},\psi_{1}),\dots,\bm{a}(\mu_{R},\psi_{R})]\in\mathbb{C}^{M\times R}$, the receive array response matrix $\bm{A}_{R}=[\bm{a}(\rho_{1}),\dots,\bm{a}(\rho_{R})]\in\mathbb{C}^{Q\times R}$, and the diagonal path gain matrix $\bm{D}=\text{diag}([\sqrt{P_{T}}\alpha_{1},\dots,\sqrt{P_{T}}\alpha_{R}])\in\mathbb{C}^{R\times R}$. Using the Khatri-Rao product ($\diamond$), the received signal for the $m$-th beam is reformulated as:
\begin{equation}
    \bm{Y}_{m} = \left[ (\bm{f}_{m}^{H}(\delta_{p},\nu_{q})\bm{A}_{T}) \diamond (\bm{W}^{H}\bm{A}_{R}\bm{D}) \right] {\bm{C}} + \bm{N}_{m}.
\end{equation}

By probing the channel using all $M$ transmit beamforming vectors, we construct the global observation matrix $\bm{Y} \in \mathbb{C}^{M \times QL}$ by stacking the row-vectorized matrices $\bm{Y}_{m}$ along the beam dimension, i.e.,
\begin{equation}
    \bm{Y}=
    \begin{bmatrix}
        \mathrm{vec}^{T}(\bm{Y}_{1})\\
        \mathrm{vec}^{T}(\bm{Y}_{2})\\
        \vdots\\
        \mathrm{vec}^{T}(\bm{Y}_{M})
    \end{bmatrix}
    \in\mathbb{C}^{M\times QL},
\end{equation}
where $\mathrm{vec}(\cdot)$ stacks the columns of its matrix argument. This yields
\begin{equation}
    \label{eq:padrao}
    \bm{Y} = \bm{F}^{H}\bm{A}_{T}\left({\bm{C}}^{T}\diamond \bm{W}^{H}\bm{A}_{R}\bm{D}\right)^{T} + \bm{N} \in \mathbb{C}^{M\times QL},
\end{equation}
where $\bm{F}=[\bm{f}_{1}(\delta_{p},\nu_{q}),\dots,\bm{f}_{M}(\delta_{p},\nu_{q})]\in\mathbb{C}^{M\times M}$ collects the probing beamforming vectors and $\bm{N}\in\mathbb{C}^{M\times QL}$ is the noise matrix. This compact multilinear structure captures all system dimensions and naturally paves the way for the tensor formulation.

\section{Hardware Constrained PARAFAC (HCP) based Channel Estimation}

The received signal \eqref{eq:padrao} model exhibits a low-rank multilinear structure across the transmit-beam, receive-beam, and pilot domains. By stacking the observation matrices $\bm{Y}_m$, a third-order tensor $\bm{\mathcal{Y}}$ is obtained.
\begin{equation}
    \bm{\mathcal{Y}} = \mathcal{I}_{3,R} \times_1 {\bm{A}}_{T} \times_2 {\bm{A}}_{R} \times_3 {\bm{C}}^{T} + \mathcal{N} \in \mathbb{C}^{M\times Q\times L},
    \label{eq:parafac_tensor_mode}
\end{equation}
where $\mathcal{I}_{3,R} \in \mathbb{R}^{R \times R \times R}$ is the core identity tensor, $\mathcal{N} \in \mathbb{C}^{M \times Q \times L}$ is the additive noise tensor, ${\bm{A}}_{T} = \bm{F}^{H}\tilde{\bm{A}}_{T} \in \mathbb{C}^{M \times R}$ and ${\bm{A}}_{R} = \bm{W}^{H}\tilde{\bm{A}}_{R}\bm{D} \in \mathbb{C}^{Q \times R}$ represent the effective beamspace transmit and receive channel signatures, respectively. The standard mode-1 and mode-2 matrix unfoldings of $\bm{\mathcal{Y}}$ are analytically derived as:
\begin{align}
    [\bm{\mathcal{Y}}]_{(1)} &= {\bm{A}}_{T}[{\bm{C}}^{T}\diamond{\bm{A}}_{R}]^{T}+\bm{N}_{(1)} \in\mathbb{C}^{M\times QL}, \label{eq:mode1_unfolding_new}\\
    [\bm{\mathcal{Y}}]_{(2)} &= {\bm{A}}_{R}[{\bm{C}}^{T}\diamond{\bm{A}}_{T}]^{T}+\bm{N}_{(2)} \in\mathbb{C}^{Q\times ML}, \label{eq:mode2_unfolding_new}
\end{align}
where $\diamond$ denotes the Khatri--Rao product, and $\bm{N}_{(n)}$ represents the unfolding of the noise tensor $\mathcal{N}$ along the $n$-th mode.

Conventional PARAFAC-based estimation relies on unconstrained trilinear Alternating Least Squares (ALS), which incurs high computational complexity, slow convergence, and scaling/permutation ambiguities across all tensor modes \cite{sidiropoulos2017}. In contrast, since the third-mode factor matrix is deterministically known as $\bm{C}^{T}$ from the pilot design, the trilinear problem reduces to a constrained bilinear optimization involving only the effective transmit and receive beamspace factors. This joint non-convex optimization minimizes the multidimensional reconstruction error, formulated as:
\begin{equation}
\label{eq:bals_optimization_new}
    \{\hat{{\bm{A}}}_{T}, \hat{{\bm{A}}}_{R}\} = \arg\min_{{\bm{A}}_{T}, {\bm{A}}_{R}} \left\| \bm{\mathcal{Y}} - \mathcal{I}_{3,R} \times_1 {\bm{A}}_{T} \times_2 {\bm{A}}_{R} \times_3 {\bm{C}}^{T} \right\|_{F}^{2},
\end{equation}
where the hat notation denotes their respective estimates and the factor matrices correspond to effective beamspace-domain representations incorporating the DFT beamforming operations.

To solve the global minimization problem in \eqref{eq:bals_optimization_new} efficiently, we decouple the joint optimization into alternating bilinear subproblems. First, by rewriting the objective function in terms of the mode-1 tensor unfolding, the transmit factor estimate $\hat{\bm{A}}_{T}$, is formulated as:
\begin{equation}
    \hat{{\bm{A}}}_{T} = \arg\min_{{\bm{A}}_{T}} \left\| [\bm{\mathcal{Y}]}_{(1)} - {\bm{A}}_{T} \left( {\bm{C}}^{T} \diamond {\bm{A}}_{R} \right)^{T} \right\|_F^2.
\end{equation}

Since this subproblem is linear and convex with respect to ${\bm{A}}_{T}$, it yields an optimal closed-form Least Squares (LS) solution computed via the Moore-Penrose pseudoinverse as:
\begin{equation}
    \hat{{\bm{A}}}_{T} = [\bm{\mathcal{Y}}]_{(1)} \left[ \left( {\bm{C}}^{T} \diamond {\bm{A}}_{R} \right)^{T} \right]^{\dagger}.
\end{equation}

Analogously, minimizing the reconstruction error over the mode-2 unfolding yields the receive factor estimate $\hat{{\bm{A}}}_{R}$:
\begin{equation}
    \hat{{\bm{A}}}_{R} = \arg\min_{{\bm{A}}_{R}} \left\| [\bm{\mathcal{Y}]}_{(2)} - {\bm{A}}_{R} \left( {\bm{C}}^{T} \diamond {\bm{A}}_{T} \right)^{T} \right\|_F^2.
\end{equation}
which is efficiently solved in a closed form by:
\begin{equation}
    \hat{{\bm{A}}}_{R} = [\bm{\mathcal{Y}}]_{(2)} \left[ \left( {\bm{C}}^{T} \diamond {\bm{A}}_{T} \right)^{T} \right]^{\dagger}.
\end{equation}

This alternating optimization cycle constitutes the Hardware-Constrained PARAFAC (HCP) estimation, which iteratively updates both spatial factor matrices until convergence is met.

Once $\hat{\bm{A}}_{T}$ and $\hat{\bm{A}}_{R}$ are obtained, the corresponding transmit and receive steering matrices are estimated as $\hat{\tilde{\bm{A}}}_{T} = \bm{F}\hat{\bm{A}}_{T}$ and $\hat{\tilde{\bm{A}}}_{R}\hat{\bm{D}} = \bm{W}\hat{\bm{A}}_{R}$, where $\bm{F}\bm{F}^{H} = \bm{I}_{M}$ and $\bm{W}\bm{W}^{H} = \bm{I}_{Q}$. Next, the angular parameters corresponding to the transmit and receive steering matrices are extracted using one- and two-dimensional ESPRIT algorithms, respectively, and the steering matrices $\hat{\tilde{\bm{A}}}_{T}$ and $\hat{\tilde{\bm{A}}}_{R}$ are reconstructed. Finally, the complex path-gain vector is estimated based on \eqref{eq:padrao} as $\small{\hat{\bm{\alpha}} = \left[ \bm{C}^{T} \diamond \left( (\bm{W}^{H}\hat{\tilde{\bm{A}}}_{R}) \diamond (\bm{F}^{H}\hat{\tilde{\bm{A}}}_{T}) \right) \right]^{\dagger} \operatorname{vec}({\bm{Y}})}$.

\begin{algorithm}[t]
\caption{Hardware-Constrained PARAFAC (HCP) based Channel Estimation}
\label{alg:hcp_bals}

\footnotesize

\begin{algorithmic}[1]

\Require Received tensor $\bm{\mathcal{Y}}$, known CAZAC pilot matrix ${\bm{C}}^{T}$, maximum number of iterations $i_{\max}$, and convergence threshold $\delta$

\State Randomly initialize $\hat{{\bm{A}}}_{R}$. Set $i = 0$
\While{$\|e(i)-e(i-1)\| \geq \delta$ and $i < i_{\max}$}
\State Find the LS estimate of $\hat{\bm{A}}_{T}$ as:
\begin{equation*}
    \hat{{\bm{A}}}_{T} = [\bm{\mathcal{Y}}_{(1)}] \left[ \left( {\bm{C}}^{T} \diamond {\bm{A}}_{R} \right)^{T} \right]^{\dagger}.
\end{equation*}
\State Find the LS estimate of $\hat{\bm{A}}_{R}$ as:
\begin{equation*}
    \hat{{\bm{A}}}_{R} = [\bm{\mathcal{Y}}_{(2)}] \left[ \left( {\bm{C}}^{T} \diamond {\bm{A}}_{T} \right)^{T} \right]^{\dagger}.
\end{equation*}
\State Update 
\begin{equation*}
e(i) = \left\|\bm{\mathcal{Y}} - \bm{\mathcal{Y}}(i)\right\|_{F}^{2}
\end{equation*}
\EndWhile

\State Estimating transmit and receive steering matrices
\begin{equation*}
    \hat{\tilde{\bm{A}}}_{T} = \bm{F}\hat{\bm{A}}_{T} , \qquad \hat{\tilde{\bm{A}}}_{R}\hat{\bm{D}} = \bm{W}\hat{\bm{A}}_{R}
\end{equation*}
\State Apply 2D-ESPRIT to $\hat{\tilde{\bm{A}}}_{T}$ and 1D-ESPRIT to $\hat{\tilde{\bm{A}}}_{R}\hat{\bm{D}}$.
\State After estimating the parameters reconstructing the transmit and receive steering matrices
\begin{equation*}
\hat{\tilde{\bm{A}}}_{T}=[\bm{a}(\phi_{1}^{tx},\theta_{1}^{tx}),\dots,\bm{a}(\phi_{R}^{tx},\theta_{R}^{tx})], \qquad \hat{\tilde{\bm{A}}}_{R}=[\bm{a}(\phi_{1}^{rx}),\dots,\bm{a}(\phi_{R}^{rx})]
\end{equation*}
\State Estimating complex path gains:
\begin{equation*}
\hat{\bm{\alpha}} = \left[ \bm{C}^{T} \diamond \left( (\bm{W}^{H}\hat{\tilde{\bm{A}}}_{R}) \diamond (\bm{F}^{H}\hat{\tilde{\bm{A}}}_{T}) \right) \right]^{\dagger} \operatorname{vec}({\bm{Y}})
\end{equation*}

\Return $\hat{\bm{\phi}}^{tx}$, $\hat{{\bm{\theta}}^{tx}}$, $\hat{\bm{\phi}}^{rx}$ and $\hat{\bm{\alpha}}$
\end{algorithmic}
\end{algorithm}

\section{Cramér-Rao Lower Bound }

This section derives the Cramér-Rao Lower Bound (CRLB) for the joint estimation of the channel parameters in the proposed architecture. The unknown deterministic parameter vector focuses on the spatial frequencies and the fading characteristics, the real and imaginary components of the complex path gains, denoted as $\text{Re}\{\bm{\alpha}\}$ and $\text{Im}\{\bm{\alpha}\}$, respectively, the spatial frequencies $\bm{\rho}$, $\bm{\mu}$ and $\bm{\psi}$. Thus,the complete deterministic parameter vector is rearranged in the following order:
\begin{equation}
    \bm{\eta} = \left[\text{Re}\{\bm{\alpha}\}^{T}, \text{Im}\{\bm{\alpha}\}^{T}, \bm{\rho}^{T}, \bm{\mu}^{T}, \bm{\psi}^{T}\right]^{T}.
\end{equation}

For unbiased estimators, the CRLB is given by
\begin{equation}
    \operatorname{var}(\hat{\eta}_i) \geq [\bm{F}^{-1}(\bm{\eta})]_{i,i}.
\end{equation}

Consequently, the theoretical lower bound for the standard deviation of the estimation error is given by
\begin{equation}
    \sqrt{\operatorname{CRLB}(\hat{\eta}_i)} = \sqrt{[\bm{F}^{-1}(\bm{\eta})]_{ii}}.
\end{equation}

For the considered observation model under circularly symmetric complex Gaussian noise, the entries of the FIM are computed using the trace operator as follows:
\begin{equation}
    [\bm{F}(\bm{\eta})]_{i,j} = \frac{2}{\sigma_n^2} \text{Re} \left\{ \operatorname{tr} \left( \frac{\partial \bm{Y}^H(\bm{\eta})}{\partial \eta_i} \frac{\partial \bm{Y}(\bm{\eta})}{\partial \eta_j} \right) \right\},
\end{equation}
where $\bm{Y}(\bm{\eta})$ represents the noiseless observation matrix constructed from the superposition of the $R$ multipath components.
To evaluate the theoretical benchmark against the channel normalized mean-square error (NMSE), the bounding expression for every Monte-Carlo run is compactly formulated as normalized-CRLB (NCRLB):
\begin{equation}
\label{eq:crlb_nmse_compact}
{\text{NCRLB}} = \frac{\text{Tr}\left[ \bm{F}^{-1}(\bm{\eta}) \right]}{\|\mathbf{H}\|_F^2},
\end{equation}

where $\bm{H} = \sqrt{P_{T}}\sum_{r=1}^{R}\alpha_{r}\bm{a}(\phi_{r}^{rx})\bm{a}^{T}(\phi_{r}^{tx},\theta_{r}^{tx})$ represents the complete true MIMO channel matrix.


The Fisher Information Matrix (FIM) is structured according to the parameter vector
$\bm{\eta}$. The block decomposition is provided in \eqref{eq:fim_block_matrix}.

\begin{figure*}[t]
\begin{equation}
\small
\bm{F}(\bm{\eta})
=
\frac{2}{\sigma_n^2}\text{Re}
\left[
\begin{array}{ccccc}
\mathbf{F}_{\text{Re}\{\bm{\alpha}\}\text{Re}\{\bm{\alpha}\}} & \mathbf{F}_{\text{Re}\{\bm{\alpha}\}\text{Im}\{\bm{\alpha}\}} & \mathbf{F}_{\text{Re}\{\bm{\alpha}\}\rho} & \mathbf{F}_{\text{Re}\{\bm{\alpha}\}\mu} & \mathbf{F}_{\text{Re}\{\bm{\alpha}\}\psi}
\\[0.3em]

\mathbf{F}_{\text{Re}\{\bm{\alpha}\}\text{Im}\{\bm{\alpha}\}}^{T} & \mathbf{F}_{\text{Im}\{\bm{\alpha}\}\text{Im}\{\bm{\alpha}\}} & \mathbf{F}_{\text{Im}\{\bm{\alpha}\}\rho} & \mathbf{F}_{\text{Im}\{\bm{\alpha}\}\mu} & \mathbf{F}_{\text{Im}\{\bm{\alpha}\}\psi}
\\[0.3em]

\mathbf{F}_{\text{Re}\{\bm{\alpha}\}\rho}^{T} & \mathbf{F}_{\text{Im}\{\bm{\alpha}\}\rho}^{T} & \mathbf{F}_{\rho\rho} & \mathbf{F}_{\rho\mu} & \mathbf{F}_{\rho\psi}
\\[0.3em]

\mathbf{F}_{\text{Re}\{\bm{\alpha}\}\mu}^{T} & \mathbf{F}_{\text{Im}\{\bm{\alpha}\}\mu}^{T} & \mathbf{F}_{\rho\mu}^{T} & \mathbf{F}_{\mu\mu} & \mathbf{F}_{\mu\psi}
\\[0.3em]

\mathbf{F}_{\text{Re}\{\bm{\alpha}\}\psi}^{T} & \mathbf{F}_{\text{Im}\{\bm{\alpha}\}\psi}^{T} & \mathbf{F}_{\rho\psi}^{T} & \mathbf{F}_{\mu\psi}^{T} & \mathbf{F}_{\psi\psi}
\end{array}
\right],
\label{eq:fim_block_matrix}
\end{equation}
\hrulefill
\end{figure*}

\section{Simulation Results}

This section evaluates the proposed Hardware-Constrained PARAFAC (HCP) estimator against a conventional least-squares (LS) baseline and the derived analytical CRLBs. The simulated mmWave MIMO system follows Section~\ref{sec:sectionTwo}, with parameters summarized in Table~\ref{tab:parameters}.

\begin{table}[t]
\centering
\caption{Simulation parameters.}
\label{tab:parameters}
\renewcommand{\arraystretch}{1.2}
\begin{tabular}{@{}llc@{}}
\hline
\textbf{Parameter} & \textbf{Description} & \textbf{Value} \\ \hline
$M_h \times M_v$   & Transmitter URA size                  & $4 \times 4$ \\
$Q$                & Receiver ULA elements              & $16$ \\
$L$                & CAZAC sequence length        & $16$ \\
$R$                & Propagation paths            & $3$ \\
$K$                & Monte Carlo realizations     & $1000$ \\ \hline
\end{tabular}
\end{table}

Performance is averaged over $K$ Monte Carlo trials across varying signal-to-noise ratios $\text{SNR}=P_T/\sigma_n^2$. Channel reconstruction accuracy is measured via the empirical Normalized Mean Square Error (NMSE):
\begin{equation}
    \text{NMSE}(\bm{\hat{H}}) = \frac{1}{K}\sum_{k=1}^{K} \frac{\|\mathbf{H}^{(k)} - \hat{\mathbf{H}}^{(k)}\|_F^2}{\|\mathbf{H}^{(k)}\|_F^2},
\end{equation}
where $\mathbf{H}^{(k)}$ and $\hat{\mathbf{H}}^{(k)}$ represent the true and estimated channel matrices at the $k$-th run, respectively. The parameter estimation accuracy is quantified by the Root Mean Square Error (RMSE):
\begin{equation}
    \text{RMSE}(x) = \sqrt{\mathbb{E} \left[ |x - \hat{x}|^2 \right]}, \quad x \in \{\phi^{\text{tx}}, \theta^{\text{tx}}, \phi^{\text{rx}}, \alpha\},
\end{equation}
where $x$ represents the true parameter  and $\hat{x}$ denotes its corresponding estimate.

\begin{figure}[t]
    \centering
    \includegraphics[width=0.80\linewidth]{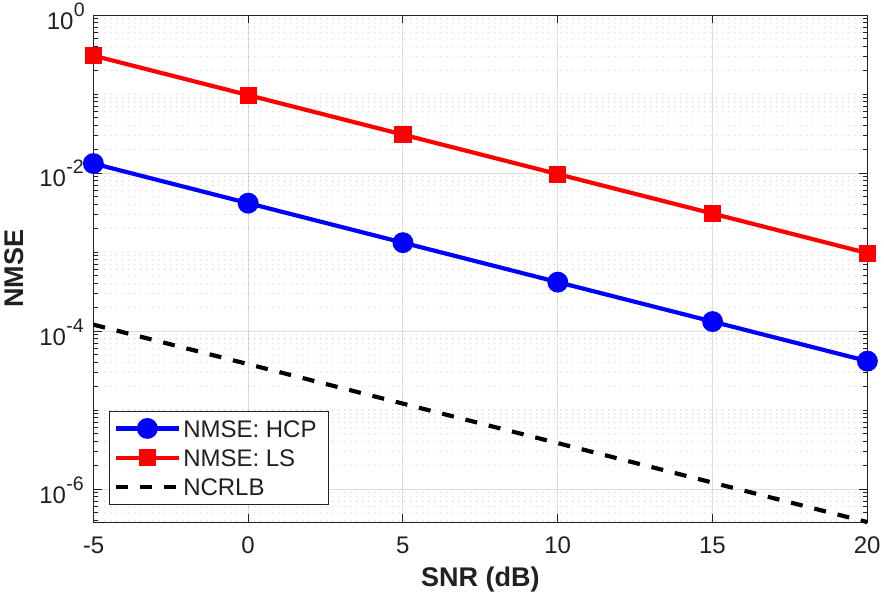}
    \caption{NMSE of the reconstructed complete MIMO channel matrix $\mathbf{H}$ versus SNR.}
    \label{fig:nmse}
\end{figure}
Figure~\ref{fig:nmse} compares the channel reconstruction NMSE of the proposed HCP estimator against the conventional LS benchmark \cite{kay1993}. The proposed framework consistently outperforms LS across the entire SNR range by explicitly enforcing the low-rank multilinear coupling among spatial signatures, path gains, and CAZAC pilots. Unlike the LS approach, which treats the channel as an unstructured matrix, this multilinear constraint drastically reduces the effective degrees of freedom, thereby enhancing robustness to noise and multipath interference. Furthermore, the proposed NMSE tightly approaches the theoretical NCRLB at moderate-to-high SNRs, validating its asymptotic efficiency and confirming that the tensor model fully captures the underlying data structure in the BM-assisted scenario.

\begin{figure}[t]
    \centering
    \includegraphics[width=0.80\linewidth]{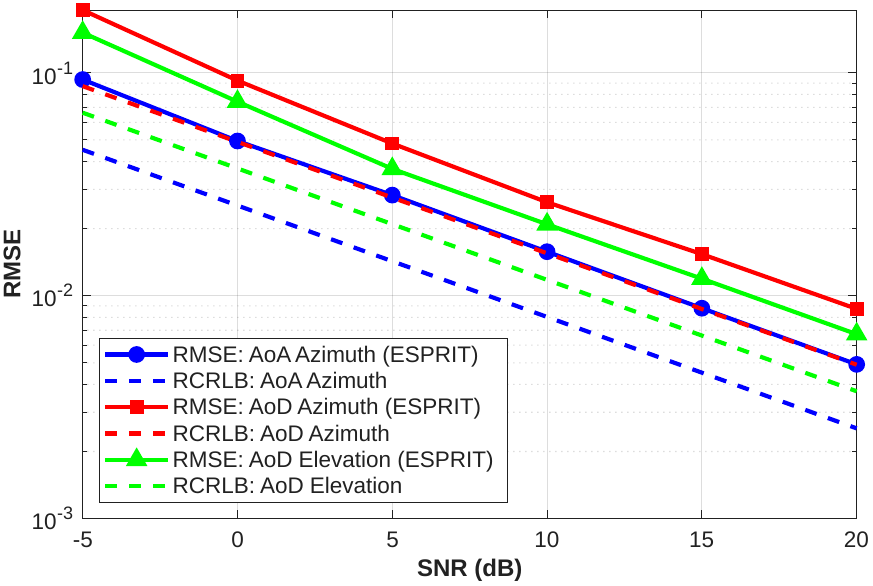}
    \caption{RMSE of AoA/AoD estimation compared against the theoretical CRLB.}
    \label{fig:rmse}
\end{figure}

Figure~\ref{fig:rmse} evaluates the RMSE of the receive AoA and transmit azimuth/elevation AoDs, which are extracted from the converged HCP spatial factors via 1D and 2D ESPRIT. At low SNRs, the estimates exhibit the classical threshold effect of subspace methods, as noise perturbations degrade the rotational-invariance relationships, creating an expected gap relative to the square-root-CRLB (RCRLB). However, as the SNR increases, the RMSE curves rapidly converge to their respective CRLBs. This asymptotic efficiency demonstrates that the proposed tensor factorization provides highly accurate spatial subspaces for high-resolution estimation. Crucially, the 2D-ESPRIT step inherently preserves the azimuth-elevation pairing via joint diagonalization, thereby bypassing any explicit parameter-association overhead. Ultimately, these results confirm the framework's capability to reliably extract the physical parameters essential for mmWave beam alignment.

\begin{figure}[t]
    \centering
    \includegraphics[width=0.80\linewidth]{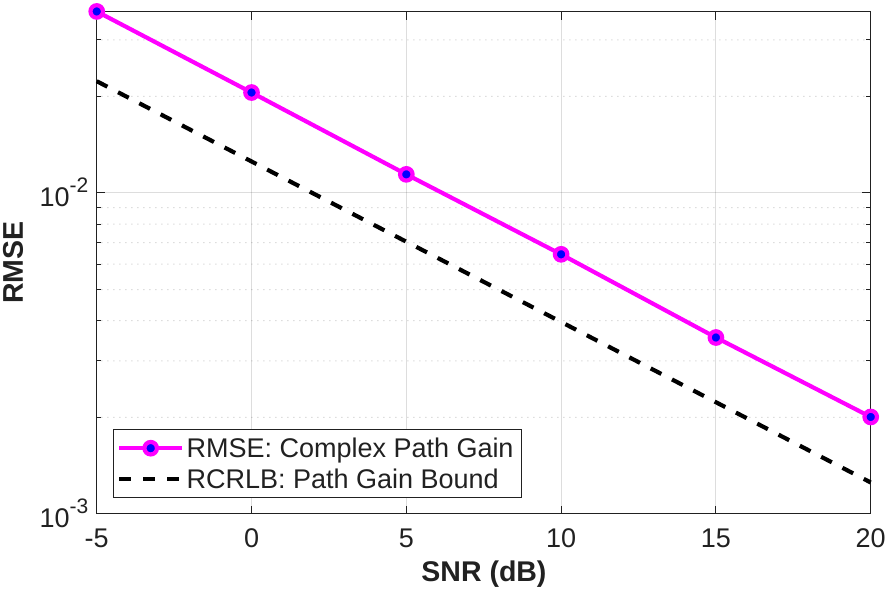}
    \caption{RMSE of the estimated complex path gains (matrix $\mathbf{D}$) versus SNR.}
    \label{fig:gains}
\end{figure}

Figure~\ref{fig:gains} evaluates the complex path-gain RMSE against the corresponding CRLB. Because path gains are tightly coupled with both spatial factors, their extraction is highly sensitive to inaccuracies in the estimated steering responses. Nonetheless, the proposed estimator tightly tracks the theoretical bound at moderate-to-high SNRs. This convergence confirms that the framework accurately preserves both path amplitude and phase information, effectively resolving the inherent PARAFAC scaling ambiguities via the deterministic pilot structure and subsequent least-squares refinement. Overall, the numerical results validate that the proposed tensor-based method achieves joint, high-accuracy recovery of spatial signatures and path gains, ensuring robust channel reconstruction while maintaining the low hardware complexity of BM-based beamforming.

\section{Conclusions}
This paper proposed a low-complexity tensor-structured HCP framework for channel parameter estimation in Butler-matrix-assisted mmWave MIMO systems. By exploiting the multilinear signal structure and known CAZAC pilots, the method efficiently recovers spatial factors, enabling high-resolution AoA/AoD extraction via 1D and 2D ESPRIT with automatic parameter pairing. The CRLB was also derived for the considered BM-assisted tensor-structured estimation scenario, providing a theoretical benchmark for the channel, angular, and path-gain estimates. Numerical results demonstrate that the proposed estimator significantly outperforms conventional least squares (LS) and asymptotically approaches the derived CRLB in the moderate- and high-SNR regimes.



\begin{thebibliography}{99}

\bibitem{elbir2022}
A. M. Elbir, K. V. Mishra, M. R. Bhavani Shankar, and B. Ottersten,
``A family of deep learning architectures for channel estimation and hybrid beamforming in multi-carrier mm-wave massive MIMO,''
\emph{IEEE Transactions on Cognitive Communications and Networking}, vol. 8, no. 2, pp. 642--656, Jun. 2022.

\bibitem{asim2023twc}
F. E. Asim, C. C. Cavalcante, F. Antreich, A. L. F. de Almeida, and J. A. Nossek,
``Efficient hybrid A/D beamforming for millimeter-wave systems using Butler matrices,''
\emph{IEEE Transactions on Wireless Communications}, vol. 22, no. 2, pp. 1001--1013, Feb. 2023.

\bibitem{asim2021twc}
F. E. Asim, F. Antreich, C. C. Cavalcante, A. L. F. de Almeida, and J. A. Nossek,
``Two-dimensional channel parameter estimation for millimeter-wave systems using Butler matrices,''
\emph{IEEE Transactions on Wireless Communications}, vol. 20, no. 4, pp. 2670--2684, Apr. 2021.

\bibitem{sidiropoulos2017}
N. D. Sidiropoulos, L. De Lathauwer, X. Fu, K. Huang, E. E. Papalexakis, and C. Faloutsos,
``Tensor decomposition for signal processing and machine learning,''
\emph{IEEE Transactions on Signal Processing},
vol. 65, no. 13, pp. 3551--3582, Jul. 2017.

\bibitem{dealmeida2023overview}
G. Favier and D. S. Rocha,
``Overview of tensor-based cooperative MIMO communication systems---Part 1: Tensor modeling,''
\emph{Entropy}, vol. 25, no. 8, Art. no. 1181, Aug. 2023.

\bibitem{liu2025nearfield}
L. Jiang, J. Guan, J. Du, W. Jiang, and Y. Cheng,
``Near-field channel parameter estimation and localization for mmWave massive MIMO-OFDM ISAC systems via tensor analysis,''
\emph{Sensors}, vol. 25, no. 16, Art. no. 5050, Aug. 2025.

\bibitem{chu1972}
D. C. Chu,
``Polyphase codes with good periodic correlation properties,''
\emph{IEEE Transactions on Information Theory},
vol. 18, no. 4, pp. 531--532, Jul. 1972.

\bibitem{andrews2022}
J. G. Andrews,
``A Primer on Zadoff Chu Sequences,''
arXiv preprint arXiv:2211.05702, 2022.

\bibitem{roy1989}
R. Roy and T. Kailath,
``ESPRIT---Estimation of signal parameters via rotational invariance techniques,''
\emph{IEEE Transactions on Acoustics, Speech, and Signal Processing},
vol. 37, no. 7, pp. 984--995, Jul. 1989.

\bibitem{kay1993}
S. M. Kay,
``Fundamentals of Statistical Signal Processing: Estimation Theory,''
Prentice-Hall, 1993.

\end{thebibliography}
\end{document}